\documentclass[10pt,twoside,BCOR7mm,DIV15,headinclude,footexclude,cleardoubleempty,idxtotoc]{scrartcl}

\usepackage[english]{babel}
\usepackage{graphicx}
\usepackage{hyperref}
\usepackage{scrpage2}
\usepackage{hyperref}
\usepackage{ifthen}

\makeatletter
\renewcommand{\@biblabel}[1]{}
\renewcommand{\@cite}[2]{%
{#1\ifthenelse{\boolean{@tempswa}}{,#2}{}}}
\makeatother

\hypersetup{breaklinks=true
,colorlinks=true,linkcolor=black,urlcolor=blue
,citecolor=black}

\ofoot{\thepage}
\ifoot{}

\setheadsepline{1pt}

\setkomafont{pagehead}{\normalfont\sffamily}
\setkomafont{pagenumber}{\normalfont\rmfamily}

\usepackage{booktabs}
\usepackage{amsmath}
\usepackage{amssymb}
\usepackage{multicol}
\usepackage{float}

\makeatletter
\newcommand{\listofcontributions}{\@starttoc{con}}

\newcommand{\l@contribution} {\@dottedtocline{1}{1.5em}{2.3em}}
\makeatother

\newenvironment{contribution}{
\setcounter{section}{0}
\setcounter{figure}{0}
\setcounter{table}{0}
\begin{flushleft}
{\em Clumping in Hot Star Winds \\
W.-R.\ Hamann, A.\ Feldmeier \& L.\ Oskinova, eds.\\
Potsdam: Univ.-Verl., 2007 \\
URN: http://nbn-resolving.de/urn:nbn:de:kobv:517-opus-13981
} 
\end{flushleft}
}{
\newpage
\lehead{}
\rohead{}
}

\begin{document}

\setlength{\baselineskip}{2.5ex}

\begin{contribution}

\lehead{N.\ Smith}

\rohead{Independent Signs of Lower Mass-Loss Rates}

\begin{center}
{\LARGE \bf Independent Signs of Lower Mass-Loss Rates for O-Type Stars}\\
\medskip

{\it\bf Nathan Smith}\\

{\it University of California, Berkeley, USA}

\begin{abstract}

I discuss observational evidence -- independent of the direct spectral
diagnostics of stellar winds themselves -- suggesting that mass-loss
rates for O stars need to be revised downward by roughly a factor of
three or more, in line with recent observed mass-loss rates for
clumped winds.  These independent constraints include the large
observed mass-loss rates in LBV eruptions, the large masses of evolved
massive stars like LBVs and WNH stars, WR stars in lower metallicity
environments, observed rotation rates of massive stars at different
metallicity, supernovae that seem to defy expectations of high
mass-loss rates in stellar evolution, and other clues.  I pay
particular attention to the role of feedback that would result from
higher mass-loss rates, driving the star to the Eddington limit too
soon, and therefore making higher rates appear highly implausible.
Some of these arguments by themselves may have more than one
interpretation, but together they paint a consistent picture that
steady line-driven winds of O-type stars have lower mass-loss rates
and are significantly clumped.

\end{abstract}
\end{center}

\begin{multicols}{2}

\section{Introduction}

Before giving a list of observational reasons to favor clumped-wind
mass-loss rates, I'll just clarify a few terms.  When I mention
``standard'' or ``unclumped'' mass-loss rates, I am referring to the
mass-loss rates derived primarily from H$\alpha$ or radio continuum
observations with the assumption of homogeneous winds (de Jager et
al.\ 1988; Nieuwenhuijzen \& de Jager 1990; NdG).  When I refer to
``lower'' or ``clumped'' mass-loss rates, I am referring to these same
mass-loss rates reduced by adopting a clumping factor.  I do not refer
to theoretical mass-loss predictions, such as those by Vink et al.\
(2001), which appear to be in line with the moderately-clumped rates.

\section{Mass Budget: LBV Eruptions}

In the evolution of very massive stars with initial masses above
$\sim$60 M$_{\odot}$, the standard mass-loss rates would have an O
star shed most of its initial mass by the end of the main sequence,
followed by a WNH or LBV line-driven wind that removes the remaining H
envelope to yield a $\lesssim$20 M$_{\odot}$ WR star.  This scenario
does not allow room for giant eruptions of LBVs, which can remove
something like 10 M$_{\odot}$ in a few years, and which seem to happen
multiple times (see Smith \& Owocki 2006).  I have discussed these
events and their role in stellar evolution ad nauseam, but the main
point here is that line-driven winds on the main sequence need to be
lower by a factor of a few in order leave enough mass on the star at
the end of core-H burning so that they can supply enough ejecta for
these outbursts.

\section{Mass Budget: High Masses of LBVs and WNH Stars}

By the same token, O-star mass-loss rates need to be lower than the
standard rates in order to agree with measured masses of stars at the
end of core-H burning.  An example that I sometimes mention is $\eta$
Carinae, because it is well studied.  We think the primary star in the
$\eta$ Car system has a present-day mass of order 100 M$_{\odot}$ (it
could be higher) if it is not violating the classical Eddington limit,
and we think it has reached the end or passed the end of core-H
burning because of the observed nitrogen enrichment in its ejecta.  It
has lost at least 20 M$_{\odot}$, perhaps 30 M$_{\odot}$, in violent
LBV eruptions in just the past few thousand years (Smith et al.\
2003), in addition to its steady wind.  That means the star made it to
the end of the main sequence with a mass of 120--130 M$_{\odot}$ still
bound to the star.  If we believe that the upper limit to the initial
mass of stars is about 150 M$_{\odot}$ (e.g., Figer 2005), then $\eta$
Car could have lost only about 20-30 M$_{\odot}$ as an O star and WNH
star combined during its first 3 Myr.  This could only be the case if
O star mass-loss rates are reduced by at least a factor of 3, probably
more.

Similar arguments apply to the Pistol star and the luminous WNH stars
(see Smith \& Conti 2008), with estimated present-day masses as high
as 80--120 M$_{\odot}$ measured in binaries.  These stars are at or
very near the end of core-H burning, having suffered mass loss from
steady line-driven winds as O stars for roughly 3 Myr as well.  With
the standard mass-loss rates, they could not exist with their
present-day masses.

\section{Feedback and High Mass-Loss}

Another problem is that the higher ``standard'' mass-loss rates lead
to an unphysical predicament -- they make the star go berzerk too
early.  I am referring to a feedback loop introduced by high mass-loss
rates (see Smith \& Conti 2008).  Namely, as mass loss reduces a
star's mass and its luminosity simultaneously climbs due to core
evolution, the star will creep closer to the Eddington limit.
However, as the Eddington factor climbs, CAK theory (Castor et al.\
1975) tells us that the mass-loss rate will climb even faster.
Eventually the star will exceed the classical Eddington limit; when
this happens depends essentially on the initial mass-loss rate.

\begin{figure}[H]
\begin{center}
\includegraphics[width=\columnwidth]{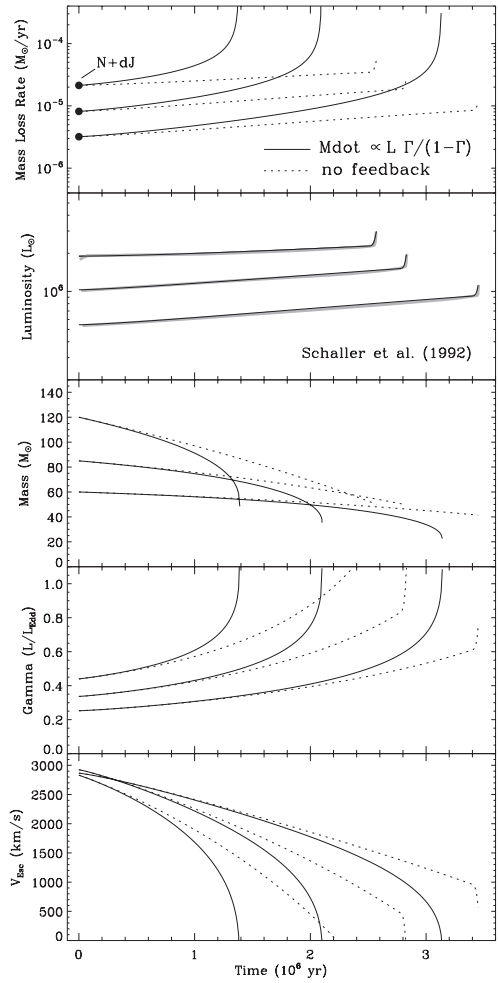}
\caption{Mass-loss rate evolution adopting standard (NdJ) initial
mass-loss rates with feedback (solid) and without (dashed).  See Smith
\& Conti (2008).\label{fig1}}
\end{center}
\end{figure}

\begin{figure}[H]
\begin{center}
\includegraphics[width=\columnwidth]{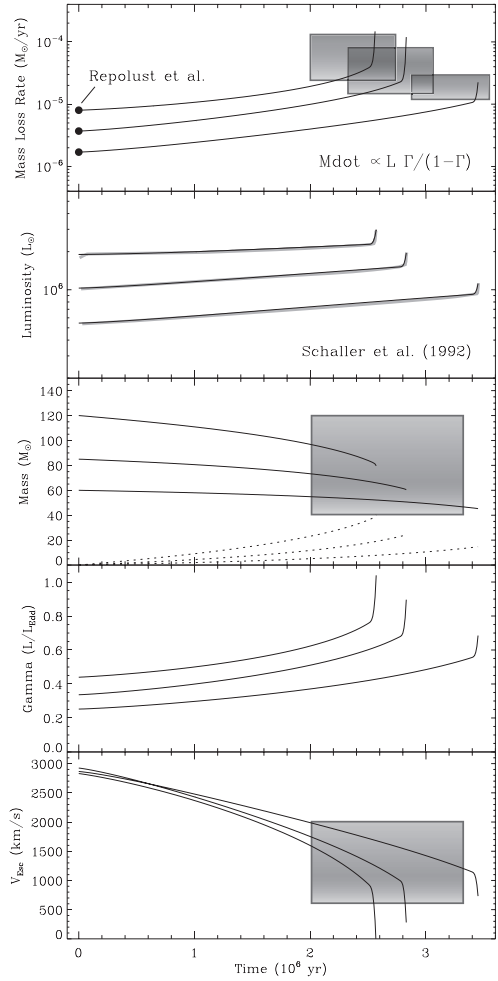}
\caption{Mass-loss rate evolution adopting moderately-clumped initial
mass-loss rates from Repolust et al.\ (2006) with feedback.  Shaded
boxes indicate observed range of properties for WNH stars; see Smith
\& Conti (2008).\label{fig1}}
\end{center}
\end{figure}

Figure 1 shows the mass-loss rate evolution predicted on the
main-sequence, taking the standard mass-loss rates as the intial
rates.  For higher masses and luminosities, the mass-loss rate
skyrockets and the star exceeds the classical Eddington limit after
only about 1 Myr.  This is too early, since almost no massive stars in
very young clusters are surrounded by LBV-type shells ($\eta$ Car is
the only one, and its age is thought to be $\sim$3 Myr).  Furthermore,
the predicted {\it change} in mass-loss rate with feedback disagrees
with the observed trend (dashed in Fig.\ 1).

By contrast, Figure 2 shows a more sensible mass-loss evolution,
starting with initial mass-loss rates for O stars that are reduced by
a factor of about 3 (rates from Repolust et al.\ 2006).  These rates,
including the expected effects of feedback, yield stellar properties
near the end of core-H burning that agree with observations of WNH
stars.  This topic is discussed in more detail by Smith \& Conti
(2008).

\section{Type IIn Supernovae}

There exists a population of supernovae that argues against strong
mass loss in steady line-driven winds as the dominant mode of
mass-loss in massive stars.  These are the Type IIn supernovae, named
for the ``narrow'' lines of H in their spectra.  There are two reasons
they contradict high mass-loss rates.

The first reason is because Type IIn supernovae are thought to mark
the deaths of very massive stars -- in some cases the {\it most}
massive stars like $\eta$ Car.  Their deaths prove that in some cases,
at roughly Solar metallicity, massive stars face death with their H
envelopes intact.  Second, the Type IIn supernovae show evidence for
huge blasts of eruptive mass loss shortly before they exploded.  These
supernova precursor events are suspiciously similar to giant LBV
eruptions (Smith \& Owocki 2006), shedding a few to 10s of M$_{\odot}$
of H-rich material in the decade before core collapse.  Thus, Type IIn
supernovae argue against the conventional wisdom that all massive
stars should shed their H envelopes via line-driven winds to form WR
stars.  See Smith et al.\ (2007) and Gal-Yam et al.\ (2007) for more
details.

\section{GRBs and WR Stars at Low-Z}

A prediction of the relatively high ``standard'' mass-loss rates is
that winds will dominate mass loss for massive stars, steadily
removing the outer layers of a star through line-driven winds until a
He-rich WR star appears.  Because these winds are metallicity
dependent, it becomes more difficult to make WR stars at low
metallicity.  This runs counter to the observations that long-duration
GRBs (associated with Type Ic supernovae that result from the deaths
of WR stars) are found only in low metallicity galaxies, and that WR
stars are seen in abundance in some low-Z galaxies, like IC~10 and
I~Zw~18.  The existence of WR stars at low Z is sometimes explained by
close binary mass transfer, but this cannot explain all of them.
Moffatt et al.\ (these proceedings) have shown that the binary
fractions among WR stars in the SMC, LMC, and Milky Way are not
significantly different (i.e., there are single WR stars or WR stars
in very wide binaries in the SMC).  This suggests that some mechanism
other than binary interaction (i.e. continuum-driven LBV eruptions;
Smith \& Owocki 2006) must be responsible for the envelope removal
that leads to WR stars in all three environments, not line-driven
winds.  One alternative way to explain the existence of WR stars at
low Z is that rapid rotation and efficient mixing leads to homogeneous
evolution (e.g., Hirschi, these proceedings), but the high initial
rotation rates required will only apply to a small fraction of stars.

\section{Z-Independent O-Star Rotation}

Mass-loss via steady line-driven winds throughout core-H burning as O
stars should lead to significant loss of angular momentum if the
``standard'' mass-loss rates apply.  Thus, there should be a clear
trend in rotation rate and metallicity among O stars.  Penny et al.\
(2004) have searched for this effect, but find no convincing
difference in the rotation period distribution of O stars in the
Galaxy, LMC, and SMC.  This, in turn, would argue strongly that the
mass and angular momentum loss is in fact much lower than given by the
standard rates, arguing that winds are indeed clumped.  The mass-loss
reduction needs to be enough that line-driven winds no longer dominate
the mass lost during a star's lifetime.

\section{Summary: Factor of 3}

For most of the observational clues that mass-loss rates are lower, a
reduction by a factor of three seems to suffice.  It could be more,
but {\it at least} a factor of three seems to be needed.  These clues
are important because they are independent of the mass-loss rates and
clumping factors derived from the analysis of O-star winds.


\end{multicols}

\end{contribution}


\end{document}